\documentclass[%
 reprint,
 superscriptaddress,
 amsmath,amssymb,
 aps,
 pra,
floatfix
]{revtex4-1}

\usepackage[paperwidth=210mm,paperheight=297mm,centering,hmargin=2.3cm,tmargin=2.8cm,bmargin=3.4cm]{geometry}
\usepackage{booktabs}
\usepackage{subfigure}
\usepackage{graphicx,epsf}
\usepackage{amsfonts}
\usepackage{amssymb}
\usepackage{amsmath}
\usepackage{overpic}
\usepackage{braket}
\usepackage{xcolor}
\usepackage{cancel}
\usepackage{fancyhdr}
\usepackage[normalem]{ulem}

\newcommand{\be}{\begin{equation}}
\newcommand{\ee}{\end{equation}}
\newcommand{\bea}{\begin{eqnarray}}
\newcommand{\eea}{\end{eqnarray}}

\def\b{\beta}

\def\th{\theta}

\def\lb{\label}
\def\pref#1{(\ref{#1})}

\newcount\bozza \bozza=1
\ifnum\bozza=1
\newdimen\shift \shift=-2truecm
\def\lb#1{%
{\label{#1}\rlap{\kern\shift{$\scriptstyle#1$}}}}
\else\def\lb#1{\label{#1}} \fi

\allowdisplaybreaks

\begin{document}

\title{Interplay of spin waves and vortices in the two-dimensional XY model at small vortex-core energy}

\author{I. Maccari}
\affiliation{Department of Theoretical Physics, The Royal Institute of Technology, Stockholm SE-10691, Sweden}
\author{N. Defenu}
\affiliation{Institute for Theoretical Physics, Heidelberg University, 69120 Heidelberg, Germany}
\affiliation{Institute for Theoretical Physics, ETH Z\"urich, Wolfgang-Pauli-Str.\ 27, 8093 Z\"urich, Switzerland}
\author{L. Benfatto}
\affiliation{Department of Physics, Sapienza University of Rome, P.le A. Moro 2, 00185 Rome, Italy}
\affiliation{Institute for Complex Systems (ISC-CNR), UOS Sapienza, P.le A. Moro 5, 00185 Rome, Italy}
\author{C. Castellani}
\affiliation{Department of Physics, Sapienza University of Rome, P.le A. Moro 2, 00185 Rome, Italy}
\affiliation{Institute for Complex Systems (ISC-CNR), UOS Sapienza, P.le A. Moro 5, 00185 Rome, Italy}
\author{T. Enss}
\affiliation{Institute for Theoretical Physics, Heidelberg University, 69120 Heidelberg, Germany}

\begin{abstract}
  The Berezinskii-Kosterlitz-Thouless (BKT) mechanism describes universal vortex unbinding in many two-dimensional systems, including the paradigmatic XY model. However, most of these systems present a complex interplay between excitations at different length scales that complicates theoretical calculations of nonuniversal thermodynamic quantities. These difficulties may be overcome by suitably modifying the initial conditions of the BKT flow equations to account for noncritical fluctuations at small length scales. In this work, we perform a systematic study of the validity and limits of this two-step approach by constructing optimised initial conditions for the BKT flow. We find that the two-step approach can accurately reproduce the results of Monte-Carlo simulations of the traditional XY model. To systematically study the interplay between vortices and spin-wave excitations, we introduce a modified XY model with increased vortex fugacity. We present large-scale Monte-Carlo simulations of the spin stiffness and vortex density for this modified XY model and show that even at large vortex fugacity, vortex unbinding is accurately described by the nonperturbative functional renormalisation group. 
\end{abstract}
\maketitle
\section{Introduction}

More than 40 years after its observation in thin $^{4}$He films~\cite{Bishop1978}, the Berezinskii-Kosterlitz-Thouless (BKT) transition \cite{berezinsky, KT, K, review} has registered in recent years an increasing interest in the field of low-dimensional strongly correlated electron systems.
In the last decade, the class of systems in which the BKT transition has been detected has become remarkably wider, including new low-dimensional experimental systems such as quasi-2D layered  superconductors~\cite{ corson_vanishing_1999, Rendeira_NatPhys07, bilbro_temporal_2011, Lemberger_PRB85_2012, Baity_LB_PRB93_2016}, exciton-polariton systems~\cite{Nitsche2014}, cold atoms in 2D harmonic traps~\cite{dalibard_2006,Murthy_PRL115_2015} and 2D electron gases~\cite{reyren_superconducting_2007, Daptary_PhysRevB.94.085104, Monteiro_PhysRevB.96.020504} at the interface between insulating oxides in artificial heterostructures. 
At the same time, interesting new issues appeared also in conventional superconducting (SC) films. For instance, the spatial inhomogeneity of the superconducting order parameter has been observed to become more pronounced near the superconducting-insulator transition (SIT)~\cite{GLemarie_PRB87_2013, Sacepe_Nat_2011}, raising the question of its effect on the BKT phase transition~\cite{Maccari_PRB96, IM_condmat, GV_LB_2019}.

Theoretically, the BKT scaling is the key to understand the universal features of a wide range of natural phenomena ranging from DNA tangling in biology to pattern formation in complex systems~\cite{Nisoli2014,Seul1991,Stariolo2012}. The hallmark of such a scaling is the expected discontinuous jump of the phase stiffness $J_s$ from a finite value $J_s(T_{BKT})$ at  the BKT temperature $T_{BKT}$ to zero right above it. According to the famous Nelson relation\cite{Nelson}, the value $J_s(T_{BKT})$  is a universal function of the critical temperature itself, i.e.,
\be
\label{jump}
J_s(T_{BKT})=\frac{2T_{BKT}}{\pi}.
\ee
Despite the apparent simplicity of this relation, the predictive power of Eq.~\pref{jump} to infer the temperature dependence of the phase stiffness and to establish the critical temperature in real microscopic systems is far from obvious. Indeed, in the original derivation based on the celebrated renormalization group (RG) BKT equations~\cite{K, Nelson}, Eq.~\pref{jump} only accounts for the role of vortex like excitations to drive the transition, including the possible renormalization effects due to bound vortex-antivortex pairs which proliferate already below $T_{BKT}$.  First, to quantify this effect one needs a precise knowledge of the vortex-core energy $\mu_v$, which controls the vortex fugacity $g=\exp(-\mu_v/k_BT)$. Indeed, for larger vortex fugacity it is easier to  generate vortex pairs at short length scales; these suppress the phase rigidity without destroying it globally and reduce $T_{BKT}$. A second issue in real systems is the presence of additional nontopological phase excitations that also deplete the phase rigidity. How these additional excitations contribute, cooperate or interfere with vortex like excitations to determine the global $T_{BKT}$ is far from understood. 

The paradigmatic example for the difficulty to estimate $T_{BKT}$ from Eq.~\pref{jump} is provided by the XY model itself, where the BKT transition was initially discussed~\cite{berezinsky, KT, K}. The XY model also admits longitudinal phase fluctuations (or spin waves in the spin language), which are effective at low temperature to deplete the stiffness linearly in temperature even while vortex like fluctuations are thermally suppressed. To deal quantitatively with this effect the approach pursued so far in the literature relied in a \emph{two-step} procedure: first one derives effective low-energy couplings for the stiffness and fugacity, which then serve as renormalized initial conditions for the BKT flow equations. Since the work of J.~Villain within the context of 2D classical planar magnets~\cite{Villain1975}, several efforts have been undertaken to derive the universal BKT scaling directly from the microscopic variables of each model, but no consistent picture has emerged despite the existence of solvable models that display BKT scaling~\cite{Lieb1967}. To account for the nontopological phase modes of the system, in addition to the RG approach, different theoretical techniques have been proposed,  such as the self-consistent harmonic approximation~\cite{Villain1975,Pires1997}, classical Monte Carlo simulations~\cite{Prokofev2001}, Gaussian fluctuations~\cite{Bighin2016, Karle2019} and the functional renormalization group~\cite{Defenu2017, Krieg2017}.

A second example of the relevance of nonuniversal effects is the application to thin films of superconductors. In this case, it has been suggested~\cite{Kamlapure2010, mondal_phase_2011, benfatto_kosterlitz-thouless_2007, Mondal_LB_PRL107, yong_robustness_2013} that the two-step procedure should include the effect of quasiparticle excitations across the superconducting gap on the stiffness, which represent the most relevant source of depletion of the superfluid density in a superconductor. However, even including this effect the comparison with experiments has shown that the  vortex-core energy $\mu_v$ is significantly smaller than expected within the XY model. Indeed, within a microscopic BCS picture for superconductors $\mu_v$ is expected not to scale with the superfluid stiffness, but rather with the Cooper-pair condensation energy $E_c$. However, within the XY model, the ratio $\mu_v/J_s$ has been estimated~\cite{KT} long ago to be a constant $\mu_v/J_s= \pi^2/2$ that depends only on the lattice structure, as expected for a model with a single coupling constant. Once more, a quantitative check of the accuracy of such an estimate for the XY model is still lacking. More broadly, the case of superconducting films calls for a deeper understanding of the two-step RG approach for larger values of the vortex fugacity. In particular, the interplay between transverse and longitudinal phase fluctuations is expected to become more relevant as the lowering of $\mu_v$ favours thermal vortex like excitations at short length scales.

What emerges clearly from these examples is that despite the universality of the Nelson criterion Eq. \pref{jump}, the critical temperature $T_{BKT}$ itself is not universal, and it depends on microscopic details which must be properly included in the RG flow to correctly reproduce the temperature dependence of the phase stiffness. In this paper, we will systematically address this issue within the paradigm of a generalized 2D XY model with tuneable vortex-core energy. Our aim is to accurately test the two-step procedure by comparing  the numerical results from Monte Carlo (MC) simulations with the RG two-step procedure for increasing values of the vortex fugacity. To this end, we first extract the stiffness and vortex-core energy from low-temperature MC data and then use them as initial conditions for the subsequent RG flow. This procedure turns out to be rather successful for the standard XY model, with an excellent estimate of $T_{BKT}$, despite some small discrepancy in the temperature evolution of the superfluid stiffness near the transition. For increasing vortex fugacity the accuracy of the original RG equations decreases, but a good estimate of $T_{BKT}$ can still be obtained by considering either higher-order terms in the vortex fugacity~\cite{Amit}, via the self-consistent approach suggested by Timm~\cite{timm_1996} or via the nonperturbative functional renormalization group (FRG). We find that FRG provides an excellent estimate of $T_{BKT}$, even though it cannot fully capture the temperature dependence of $J_s(T)$ obtained from Monte Carlo simulations. This discrepancy can be an indication of the interplay between longitudinal and transverse phase fluctuations, which cannot be captured by the two-step RG approach.

The paper is organized as follows. In Sec.\,II we introduce the two-step procedure for the classical XY model, providing a new estimate of the vortex-core energy $\mu_v$ based on Monte Carlo numerical results. In Sec.\,III we introduce the modified 2D XY model studied in this work and we present the MC results. In Sec.\,IV we study the modified XY model via a two-step procedure, where an effective value for both the vortex-core energy $\mu_v^\text{eff}$ and the superfluid stiffness $J_{\mathrm{eff}}$ are inserted into the BKT flow equations. Finally, we conclude in Sec.\,V.

\section{The vortex-core energy of the XY model}
\label{secii}
Ever since the seminal papers by Berezinskii, Kosterlitz and Thouless \cite{berezinsky, KT, K}, the classical 2D XY model has been the reference model for the study of the BKT phase transition. The model describes the ferromagnetic nearest-neighbor interaction between planar spins with fixed modulus ($|\vec{S_i}|=1$), interacting via a ferromagnetic spin-spin coupling constant $J>0$:
\begin{align}
H_{XY}&=-J\sum_{i, \nu= \hat{x}, \hat{y}} \vec{S}_i \cdot \vec{S}_{i+\nu}\nonumber\\
&= -J\sum_{i, \nu= \hat{x}, \hat{y}} \cos(\th_i -\th_{i+\nu}),
\label{XY}
\end{align}
where $\th_i$ is the angle the $i$th spin forms with a given direction. Within the context of SC films, one can map $\vec{S}_i \to |\Delta| e^{i \theta_i}$ so that Eq.~\eqref{XY} describes only phase fluctuations. By retaining leading terms in the phase gradient one can also identify $J$ with the phase stiffness of the SC.

The standard way to treat the XY model analytically, see Ref.~\cite{review} for a review, can be sketched in two main steps:
\begin{enumerate}
\item The mapping to the Villain model~\cite{Villain1975}, and
\item the study of the BKT renormalization group equations.
\end{enumerate}
The first step consists essentially in writing Eq.~\eqref{XY} as the sum of two decoupled contributions: an effective harmonic Hamiltonian $H_{SW}$, which accounts for the longitudinal spin-waves excitations, and a vortex Hamiltonian $H_V$,  which accounts for the transverse (topological) spin excitations. %

The spin-wave fluctuations are responsible for the lack of long-range order at any finite temperature~\cite{MW}. Their anharmonicity reduces the value of the superfluid stiffness $J_s(T)$, which is equal to the bare coupling $J$ at $T=0$, continuously in temperature without inducing yet any phase transition.

The topological vortex excitations, however, drive the BKT phase transition, whose hallmark is the universal jump of the superfluid stiffness~\cite{Nelson}.
At the critical point ($T=T_{BKT}$), the proliferation of free vortices transforms the system from a quasi-ordered phase with zero magnetization and finite phase rigidity ($J_s \neq 0$) into a disordered phase in which the system is no longer rigid ($J_s=0$).

As widely discussed in the literature, the Villain model $H_{V}$ is equivalent to the Hamiltonian of a 2D Coulomb gas, where positive and negative charges  ($q_i=\pm 1$) play the role of vortices and anti-vortices, respectively. Once restricted to the case $\sum_i q_i =0$, it reads: 
\begin{equation}
H_{V}=-\pi J \sum_{i\neq j} \ln \left(\frac{|\vec{r_{ij}}|}{a}\right) q_iq_j  + 2  \mu_v \sum_{i} \, \nu_v({\bf r}_i)\ ,
\label{Hv}
\end{equation}
where $\nu_v$ is the vortex-pairs density and $a$ the lattice spacing of the discrete model. The first term in Eq.~\eqref{Hv} describes the interaction between two charges at distance $|\vec{r}_{ij}| \gg a$ from each other, while the second term corresponds to the energetic cost of nucleating vortices at the shortest length scale with vortex-core energy $\mu_v$.
 
The effect of the transverse phase fluctuation is well captured by the BKT renormalization-group equations~\cite{berezinsky, KT, K}, rigorously derived for the Coulomb gas model Eq.~\eqref{Hv} at leading order in the vortex fugacity $g \to 0$ \cite{Jose1977, Amit,  MinnhagenRev, Gulacsi1998}. The relevant variables are the dimensionless couplings
\bea
\label{K_0}
K(0)&=& \frac{\pi J}{T},\\
\label{g_0}
g(0) &=& 2 \pi e^{-\beta \mu_{v}},
\eea
expressed in terms of the spin stiffness $J$ and the vortex fugacity $g$. 

At long distances, the values of the two couplings $K$ and $g$ are obtained from the numerical solution of the celebrated BKT flow equations\,\cite{KT, K}: 
\bea
\label{BKTeq1}
\frac{dK}{dl}&=& -K^2 g^2,\\
\label{BKTeq2}
\frac{dg}{dl} &=& (2 - K) g,
\eea
where $l=\ln(r/a)$ is the rescaled length scale. The long-distance (thermodynamic) value of the superfluid stiffness $J_s$ is thus determined by $J_s \equiv T K(l \to \infty)/ \pi$.
The above equations identify two main regimes: a low-temperature regime $K > 2$, where the vortex fugacity flows to zero $g \to 0$ while the superfluid stiffness flows to a constant value $K \to K^*$, and a high-temperature regime $K <2$, where the vortex fugacity grows $g \to \infty $  and the superfluid stiffness vanishes $K \to 0$. The BKT critical temperature $T_{BKT}$ is defined as the temperature at which the flow reaches the fixed point $K=2$, $g=0$,
\be
\label{nels_crit}
K(l \to \infty, T_{BKT})= 2  \implies \frac{\pi J_s(T_{BKT})}{T_{BKT}} = 2,
\ee
and is equivalent to the Nelson criterion~\cite{Nelson}. The renormalization of the stiffness from the initial value to a smaller $J_s= TK^*/\pi$ at $T<T_{BKT}$ depends quantitatively on the value of the vortex-core energy $\mu_v$: the larger the initial value of the vortex fugacity $g$ in Eqs.~\pref{BKTeq1} and \pref{BKTeq2}, the stronger the relative suppression of the stiffness due to bound vortex-antivortex pairs which nucleate at small length scales. 

The flow Eqs.~\eqref{BKTeq1} and \eqref{BKTeq2} account only for the effect of vortex excitations, so that any other excitation that contributes to renormalize the superfluid stiffness or the vortex-core energy in temperature must be  introduced by hand. This is the key idea behind the two-step approach, which consists in incorporating the effect of noncritical fluctuations into the \emph{effective} couplings $J_{\mathrm{eff}}$ and $\mu^{\mathrm{eff}}_{v}$, which are then used in place of the bare values $J$ and $\mu_v$ as initial conditions for the BKT flow~\cite{review, Mondal_LB_PRL107, yong_robustness_2013, Defenu2017}. While for the case of 2D superconducting films the depletion of the superfluid stiffness is mainly due to quasiparticle excitations, well accounted for by the BCS expression $J_{\mathrm{eff}}=J_s^\text{BCS}(T)$~\cite{Mondal_LB_PRL107,yong_robustness_2013}, within the XY model the effective superfluid stiffness $J_{\mathrm{eff}}(T)$ is reduced with respect to $J$ by longitudinal (spin-wave like) fluctuations already at one loop order. However, since the XY model has only a single energy scale $J$, it is natural to assume that also $\mu_v$ scales with $J_{\mathrm{eff}}(T)$. A natural ansatz for the initial values of the couplings is then
\bea
J_{\mathrm{eff}}(T)&=J-\frac{T}{4}\,,\label{Jeff}\\
\mu_{v}^{\mathrm{eff}}(T)&= \gamma J_{\mathrm{eff}}(T).
\label{mu_XY}
\eea
The estimate in Eq.~\eqref{Jeff} has been obtained by including the $(\nabla \theta)^{4}$ contribution in the low energy spin-wave expansion of the XY Hamiltonian \eqref{XY}, computed with perturbation theory around its quadratic Gaussian form \cite{IM_dis_2019}, see Ref. \cite{review} for a recent review. For the vortex-core energy, Kosterlitz and Thouless~\cite{KT} derived an estimate of $\gamma$ for a square lattice geometry,
\begin{equation}
\gamma^{KT}= \frac{\pi^{2}}{2}.
\label{gamma_KT}
\end{equation}
This estimate, however, was obtained within the 2D Coulomb gas model Eq.~\eqref{Hv}, and it has not yet been verified to what extent it corresponds to the actual value of the vortex-nucleation energy in the full XY model Eq.~\eqref{XY}. 

\begin{figure}
\centering
\includegraphics[width=\linewidth]{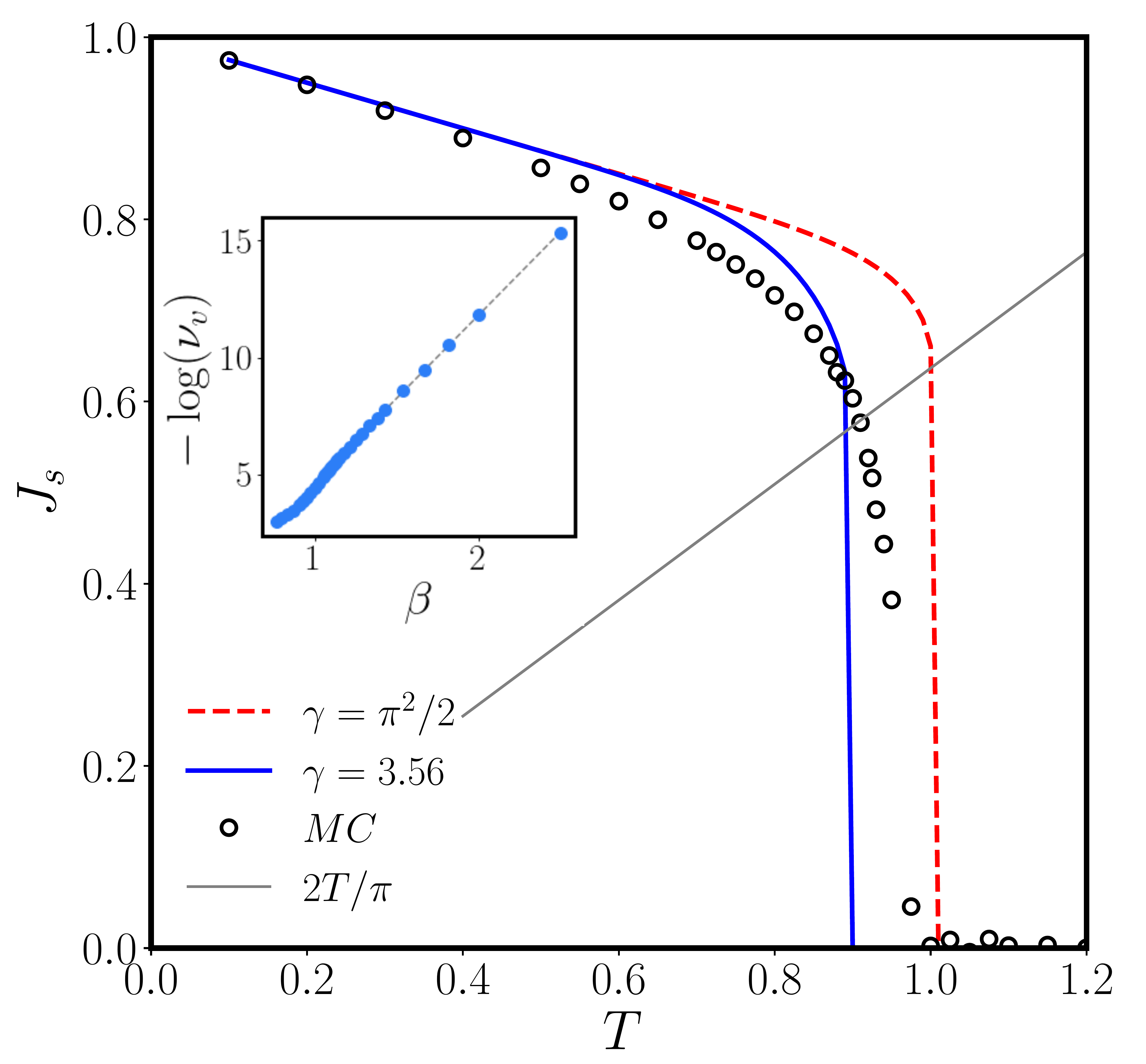}
\caption{Superfluid stiffness $J_s$ vs temperature $T$: Monte Carlo simulations for a system of linear size $L=256$ (black circles); KT estimate Eq.~\eqref{gamma_KT} for $\mu_v$ as initial condition for BKT RG equations (dashed red line); MC estimate Eq.~\eqref{gamma_MC} for $\mu_v$ as initial condition for BKT RG equations (solid blue line). The gray diagonal is the critical line where the BKT transition is expected to occur. In the inset, the MC numerical result for the vortex-pair density plotted as $-\log(\nu_v)$ (blue points) and the resulting low-temperature fit Eq.~\eqref{nu_v_fit} (dashed gray line).}
\label{Fig1}
\end{figure}
Monte Carlo simulations offer an opportunity for this verification: the superfluid stiffness $J_s$ obtained numerically from the cosine model Eq.~\eqref{XY} can indeed be compared with the RG estimate implemented via the two-step procedure.
Besides, from the numerical calculation of the vortex-pair density one can determine the vortex-core energy of the model and compare it with the Eq.~\eqref{gamma_KT}.

The superfluid stiffness $J_{s}$ is  defined as the second derivative of the free energy with respect to a phase twist $\Delta_{x}$ in a given direction,
\begin{equation}
 J_{s}^{x} \equiv -\frac{\partial^2 F(\Delta_x)}{\partial \Delta_{x}^2} \Big|_{\Delta_x=0},
 \label{Js_def}
\end{equation}
and it measures the response of the SC film to a transverse gauge field $\bf{A}$. A constant field $\bf{A}$ in a given direction, say $\mathbf{A}=A_x$, is indeed equivalent to applying a total phase twist $\Delta_x= A_xL$ across the sample of length $L$.
The superfluid stiffness
  \begin{align}
    \label{eq:Jsx}
    J_s^x = J_d^x - J_p^x
  \end{align}
  can be expressed in terms of the diamagnetic ($J_d$) and the paramagnetic ($J_p$) response functions
\begin{align}
\label{Jd}
J_d^x &= \frac{1}{L^2} \Big[ \Bigl\langle \frac{\partial^2 H}{\partial A_x^2}\Big|_{0}\Bigr\rangle  \Big], \\
\label{jp}
J_{p}^x &= \frac{\beta}{L^2} \Big[ \Bigl\langle \Bigl(\frac{\partial H}{\partial A_x}\Big|_{0} \Bigr)^2\Bigr\rangle -\Bigl\langle \frac{\partial H}{\partial A_x}\Big|_{0}  \Bigr\rangle ^2 \Big],
\end{align}
where $\langle \dots \rangle$ stands for the thermal average. The explicit expressions of $J_d$ and $J_p$ are reported in Appendix~\ref{sec:MC}.
 Apart from the superfluid stiffness, we have also measured the vortex-pair density of the system,
\begin{equation}
\nu_v =  \frac{1}{2} \rho_v= \frac{1}{2}\biggl\langle \frac{1}{L^2} \sum_{i}^{N} |v(P_i)|\biggr \rangle,
\label{nu_v}
\end{equation}
where $\rho_v$ is the vortex density and $v(P_i)=+1 (-1)$ if the plaquette $P_i$ is occupied by a (anti-)vortex and $0$ otherwise. 

Once $\nu_v(T)$ and $J_s(T)$ are determined from Monte Carlo simulations, one can extrapolate an effective 
vortex-core energy $\mu_v^{MC}(T)=\gamma^{MC} J_s(T)$ and compare it with Eq.~\eqref{gamma_KT}.  
Indeed, assuming that the system at low $T$ is formed by noninteracting vortex pairs, one can introduce the following BKT ansatz for the low-temperature pair density
\be
{\nu_v(T) =A e^{-2\b \mu^{\mathrm{eff}}_v(T)}},
\label{nu_v_fit}
\ee
where $\mu_v^{\mathrm{eff}}(T)$ is the temperature dependent vortex-core energy in Eq.\,\eqref{mu_XY}, while $A$ is a geometrical factor~\cite{timm_1996}. 
We fit the MC numerical results for $\nu_v$ to the form Eq.~\eqref{nu_v_fit}, see inset of Fig.~\ref{Fig1}, and find the parameters
\bea
\label{gamma_MC}
&\gamma^{\mathrm{MC}}= 3.56 \pm 0.01,\\
&A^{\mathrm{MC}}= 1.87 \pm 0.06.
\label{A_MC}
\eea
The numerical value of $\gamma$ displayed in Eq.~\eqref{gamma_MC} is significantly smaller than the long-standing KT estimate Eq.~\eqref{gamma_KT}.  Apparently, the effect of the spin-wave excitations and their relation to the presence of vortex pairs, which is not properly accounted for in the original derivation, is to lower the energetic cost to nucleate a vortex pair.

Then, we compare the superfluid stiffness $J^{\mathrm{MC}}_s(T)$, obtained by Monte Carlo simulations of the full XY model, with the one obtained by solving the flow Eqs.~\eqref{BKTeq1}-\eqref{BKTeq2} with the initial condition $J_{\mathrm{eff}}(T)$ and $\mu_v^{XY}(T)$ given by Eqs .~\pref{Jeff},\pref{mu_XY},\pref{gamma_KT} and \pref{gamma_MC}. The results are reported in Fig.~\ref{Fig1}:
with the KT ansatz for $\gamma$ (dashed red line in Fig.~\ref{Fig1}) the renormalized $J_{s}^{\mathrm{RG}}$ is larger than $J_s^{\mathrm{MC}}$ at higher temperatures and leads to a larger value of the critical temperature $T_{BKT}=\pi J_s(T_{BKT})/2$. On the other hand, the renormalized $J_{s}^\mathrm{RG}$ obtained using the vortex-core energy \eqref{gamma_MC} (blue line in Fig.~\ref{Fig1}) gives a very good agreement both with the MC critical temperature and with the whole temperature dependence of the superfluid stiffness. The discrepancy between the two at high temperature $T>T_{BKT}$ can be easily explained in terms of finite-size effects in the MC numerical results. However,  the deviation observed at low temperatures ($T<T_{BKT}$) could be given to the approximation which truncates high order spin-wave contributions to the depletion of the superfluid stiffness Eq. \eqref{Jeff} and neglects the interplay between longitudinal and transverse excitations.

\section{The modified XY model}

So far, we have reviewed the standard analytical treatment of the XY model, solving the BKT renormalization-group equations via a two-step procedure that we have improved with a new estimate for $\mu_v^{\mathrm{eff}}$. 
As already stressed, however, within the classical XY model there is no way of tuning the value of $\mu_v$ independently from the coupling $J$, as one would need when studying real systems.
In particular, the study of thin SC films~\cite{Mondal_LB_PRL107} has revealed a much smaller value of the vortex-core energy than expected from the XY model, a challenge for the study of the BKT transition in the limit of high vortex fugacity.

For this purpose, we introduce a modified version of the original XY model in which $\mu_v$ can be tuned independently from the spin stiffness $J$. The bare vortex-core energy in Eq.~\eqref{mu_XY} can be modified by adding an extra potential term to the Hamiltonian in Eq.~\eqref{XY}, 
\begin{align}
H_{XY}^{{\mu}}= -J\sum_{i, \nu=\hat{x},\hat{y}} \cos(\theta_i- \theta_{i+\nu})- {\mu} \sum_{i} \big( I_{P_i} \big)^2.
\label{H_2}
\end{align}
Here, $I_{P_i}$ corresponds to the spin current circulating around a single plaquette $P_i$ of area $a^2$,
\begin{equation}
\begin{split}
I_{P_i}=&\sin(\theta_i - \theta_{i+\hat{x}}) + \sin(\theta_{i+\hat{x}}- \theta_{i+\hat{x}+\hat{y}})\\ +& \sin(\theta_{i+\hat{x}+\hat{y}} - \theta_{i+\hat{y}}) + \sin(\theta_{i+\hat{y}} - \theta_{i}).
\end{split}
\label{current}
\end{equation}
The advantage of choosing an additional potential term of the form in Eqs.~\eqref{H_2}, \eqref{current} is that it is a local term, it is derivable with respect to the phase difference, and it has a direct physical interpretation in terms of currents.

For $\mu=0$ one recovers the classical XY Hamiltonian \eqref{XY} while for ${\mu}>0$ the additional potential term favours the presence of spin currents in the system. This enhances the nucleation of vortices and reduces the resulting vortex-core energy $\mu_v$. 
In the following, we will first discuss the Monte Carlo numerical study of the modified XY model. Then, we will analyse it theoretically using the RG equations for the BKT transition, implemented via the two-steps procedure.

\section{Monte Carlo simulations}

\begin{figure}[!ht]
\centering
\includegraphics[width=0.86\linewidth]{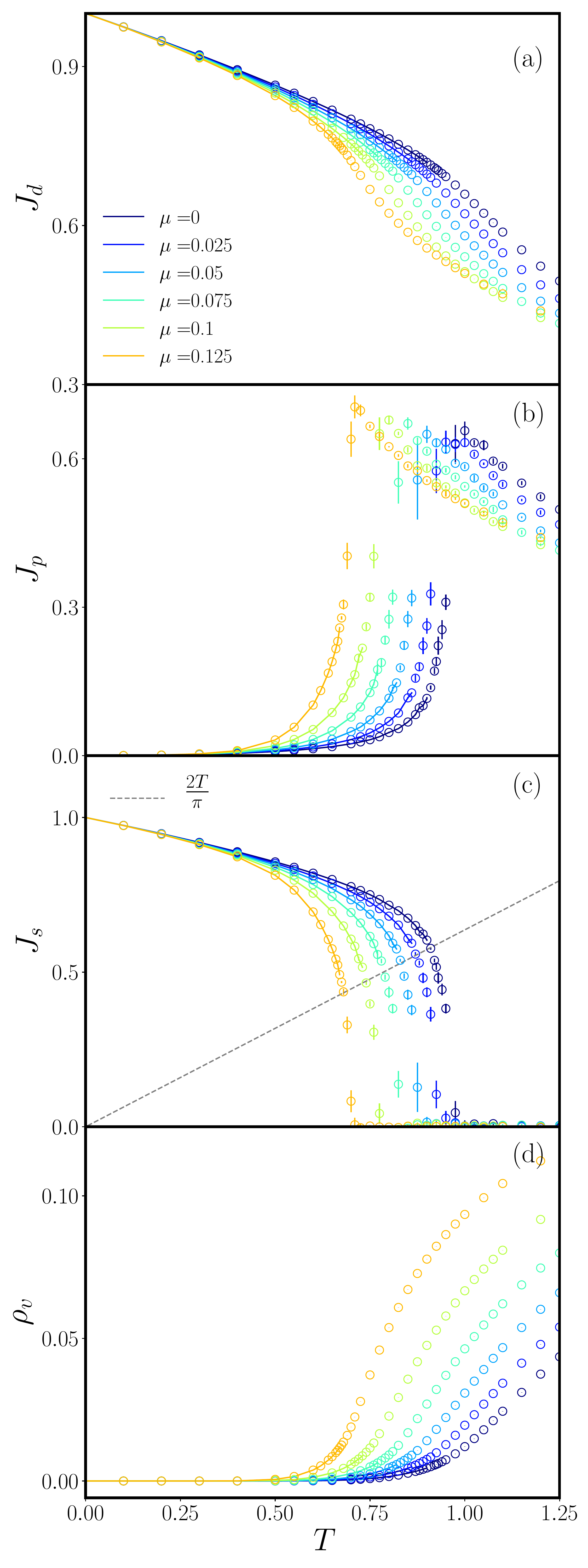}
\caption{\label{Fig3} MC analysis of the modified XY model \eqref{H_2}. Panel (a) reports the diamagnetic current $J_{d}$ vs temperature $T$. Panel (b) shows the paramagnetic currents $J_{p}$ produced by the topological excitations. Panel (c) displays the superfluid stiffness $J_{s}=J_d -J_p$, and the dashed critical line $2T/\pi$ indicates the point where the universal jump is expected to occur. Finally, panel (d) shows the vortex density $\rho_v$ as a function of the temperature.  The solid lines are guides to the eye and terminate at the BKT transition point. All data are for linear system size $L=256$ and in units of $J$.}
\end{figure}

In Fig.~\ref{Fig3}, the MC numerical results for the temperature dependence of $J_d$, $J_p$ and $J_s$ are reported for different values of $\mu$. For larger values of $\mu$ (smaller $\mu_v$) the BKT critical temperature decreases, since at lower values of the vortex-core energy the entropic term in the free energy of a single vortex $F_v= E_v -TS_v$ dominates already for lower temperature and brings forward the vortex-antivortex unbinding.

As argued above, the decoupling between transverse and longitudinal phase fluctuations cannot be rigorously applied to the XY model as it only holds for its quadratic approximation, the Villain model~\cite{Villain1975}.
At low temperatures the curves for the diamagnetic response function $J_d$ collapse onto each other and thus demonstrate the decoupling between spin-wave and vortex excitations in the $T\to 0$ limit. At higher temperatures the curves separate, and for larger $\mu$ there is a stronger depletion of the diamagnetic currents due to the interplay between spin-wave excitations and vortices, caused by the anharmonic terms in the XY coupling, see Fig.~\ref{Fig3}(a). The paramagnetic response functions in Fig.~\ref{Fig3}(b) display a very slow increase in the low-temperature regime as they only receive contributions from high-order powers in the phase gradient~\cite{IM_dis_2019}.

The temperature dependence of the superfluid stiffness in Fig.~\ref{Fig3}(c) reveals another interesting feature for increasing values of $\mu$. Indeed, for smaller values of the vortex-core energy $\mu_v$ it becomes more and more difficult to distinguish between the BKT universal jump of $J_s$ at $T=T_{BKT}^+$ and its rapid downturn, which starts already at a lower temperature. The $2T/\pi$ critical line for $\mu=0.125$ appears, in fact, to be halfway through the jump, rather than at its onset.
This aspect can be very important for a correct interpretation of experimental data, as the observation of a BKT critical line halfway across the jump can be considered an indication for a low vortex-core energy within the system.

At small vortex-core energies, the energy-entropy balance suggests the presence of a critical value $\mu=\mu^c$, above which unbound free vortices are energetically favoured even at zero temperature.  As a consequence, for $\mu> \mu^c$ the ground state of the system will change from the vortex-vacuum state with $\rho_v(T\to0)=0$ to a vortex-antivortex square-lattice crystal with $\rho_v(T\to0)=1$. A similar effect has been discussed for the 2D Coulomb gas in Refs.~\cite{LeeTeitel2, LeeTeitel, LidmarWallin96}. For the model \eqref{H_2} we find the critical value $\mu^c \simeq 0.15$ and focus henceforth on the regime $\mu< \mu^c$. A preliminary study of the vortex crystal phase $\mu>\mu_{c}$ is presented in App.~\ref{app:vlattice}, while a more complete analysis is deferred to future investigations.

\section{RG study of the modified XY model}

Let us proceed with the study of the generalised XY model \eqref{H_2} by means of the two-step RG approach.
As in Sec.~\ref{secii}, one can treat the modified XY Hamiltonian \eqref{H_2} in the spirit of the Villain approximation by separating the spin-wave from the topological excitations. It is easy to verify that $H_{SW}$ remains unchanged, while $H_v$ acquires a new term proportional to $\mu$.
As a consequence, while the low-temperature estimate for the superfluid stiffness $J_{\mathrm{eff}}$ \eqref{Jeff} remains valid for $\mu \neq 0$, one needs to identify a new renormalized value for $\mu^{\mathrm{eff}}_{v}$.

\begin{figure}
\centering
\includegraphics[width=\linewidth]{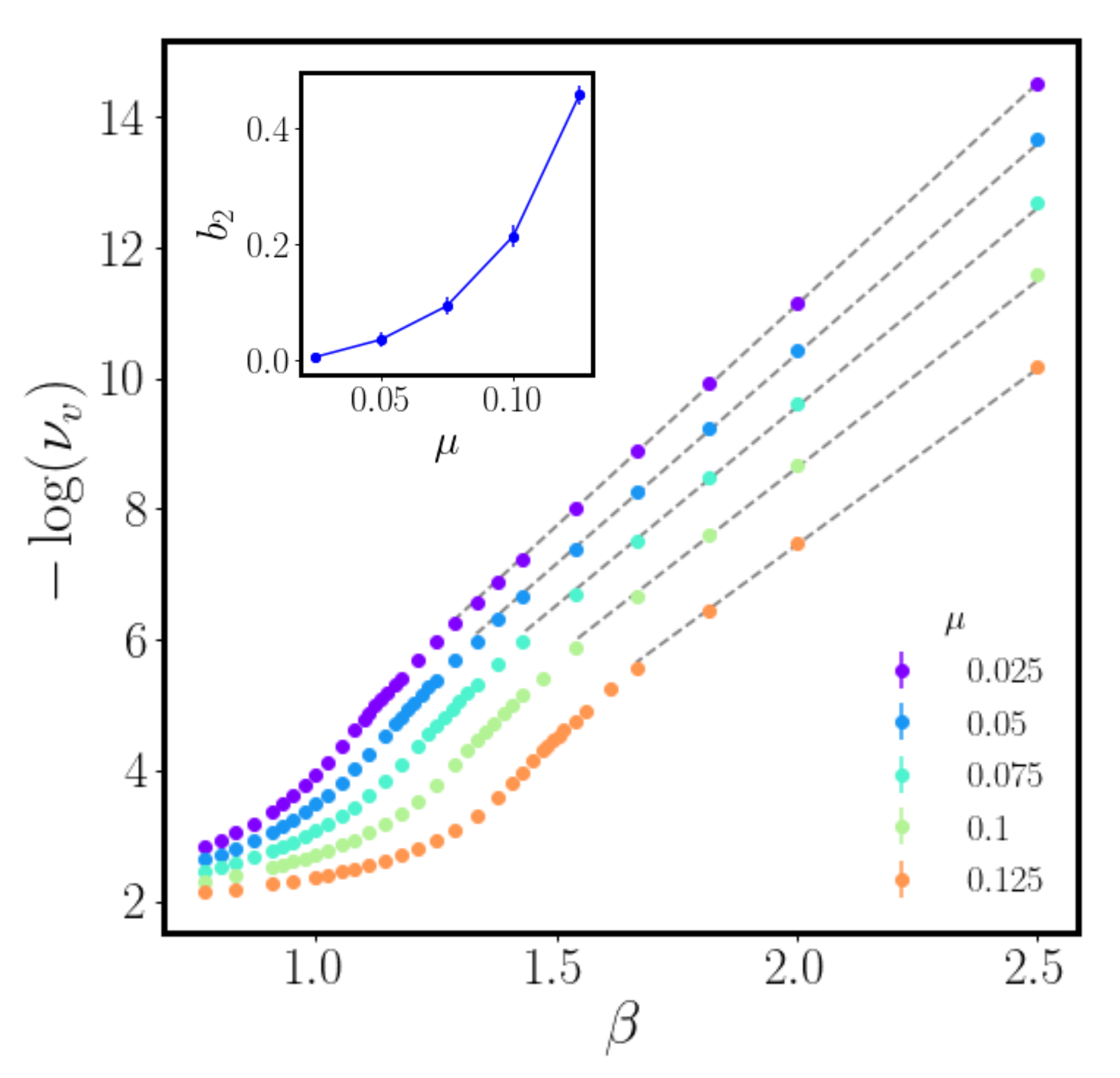}
\caption{Monte Carlo numerical results for the vortex-pair density plotted as $-\log(\nu_v)$ for all the values of $\mu$ considered. The resulting low-temperature fits are shown as dashed gray lines.  The best fitting parameters found are $b_1=6.9$, while $b_2(\mu$) is shown in the inset.}
\label{Fig4}
\end{figure}

We will proceed following the same route as before, i.e., by fitting the vortex-pair density $\nu_v$ from MC numerical simulations of the modified XY model with Eq.~\eqref{nu_v_fit}.
In this case, however, the vortex-core energy will depend on both $J$ and $\mu$.
 Moreover, as for the superfluid stiffness $J^\text{eff}_s$ \eqref{Jeff} the presence of low-temperature spin-wave fluctuations will modify the effective value of $\mu$ as well, whose temperature dependence can be considered in first approximation to be simply linear.
In light of these considerations, we have used the following ansatz for the vortex-core energy at finite $\mu$:
\be
\mu_v^\text{eff}(T,\mu) = \gamma^{\mathrm{MC}}J_{\mathrm{eff}}(T) - b_1 \mu +b_2(\mu)T.
\label{mueff1}
\ee
For the numerical fits of $\nu_v(T)$, we have fixed the values of $A$ and $\gamma^{\mathrm{MC}}$ in Eqs.~\eqref{gamma_MC}-\eqref{A_MC} and determined the best fits $b_1$ and $b_2(\mu)$.
As shown in Fig.\ref{Fig4}, they yield very good agreement with the Monte Carlo data for all values of $\mu$.

\begin{table}[ht!]
  \centering
  \begin{tabular}{ccccc}
\toprule
 $\mu$  & $\gamma^{\mathrm{MC}}$ & $b_{1}$ &     $b_{2}$ \\
\midrule
 0.000 & 3.56(1) & --- &  ---\\
 0.025 & 3.56(1) & 6.9 & 0.005(6) \\
 0.050 & 3.56(1) & 6.9 & 0.036(11) \\
 0.075 & 3.56(1) & 6.9 & 0.094(15) \\
 0.100 & 3.56(1) & 6.9 & 0.21(2) \\
 0.125 & 3.56(1) & 6.9 & 0.45(2) \\
\bottomrule
  \end{tabular}
  \caption{Numerical coefficients for the effective vortex core energy using Eq.~\eqref{mueff1}, based on the analysis of the MC data described in Fig.~\ref{Fig4}.}
  \label{tab1}
\end{table}

To assess the validity of the two-step technique for the generalised XY Hamiltonian Eq. \eqref{H_2}, one must carefully consider the two underlying approximations:
\begin{enumerate}
\item[(i)] the assumption of decoupling between topological configurations and noncritical excitations;
\item[(ii)] the BKT flow equation that is perturbative in the vortex fugacity.
\end{enumerate}
The hypothesis (i) implies that critical and noncritical fluctuations occur on well-separated scales, the firsts occurring only on short-distances. The noncritical fluctuations are indeed present at all length scales and they contribute to renormalize the couplings in play even at long distances. 
However, to properly include these contributions, they should be inserted within the RG flow equations themselves. Something that, as we discussed, is not feasible so far. 
However, the limitation to small vortex fugacity can be easily circumvented by including additional terms in the perturbative expansion or by employing a self-consistent RG scheme~\cite{MinnhagenRev}.

\subsection{The critical temperature}

Both assumptions (i) and (ii) are satisfied for the Villain model, and the computation of the critical temperature from the BKT flow equations with $J_{\mathrm{eff}}=J$ and $\mu^{\mathrm{eff}}_{v}=\mu_{v}=J\pi^{2}/2$ perfectly reproduces high-precision MC results~\cite{Janke1993}. In the XY model case with $\mu=0$ one can efficiently compute the renormalization of the effective superfluid stiffness due to spin waves via functional RG, mean-field or low-temperature expansion, yielding estimates for the BKT critical temperature \cite{Jakubczyk2016,Defenu2017,LBGiamarchiCastellani} within 5\% of the numerically exact MC value~\cite{gupta_phase_1988, gupta_critical_1992, Hasenbusch2005}. Finally, the computation of the effective superfluid stiffness for 2D Fermi gases via Landau's quasiparticle-excitation formula produces a consistent picture for the dependence of the BKT temperature on the scattering length~\cite{Bighin2016}. 

\begin{figure}
\centering
\includegraphics[width=\linewidth]{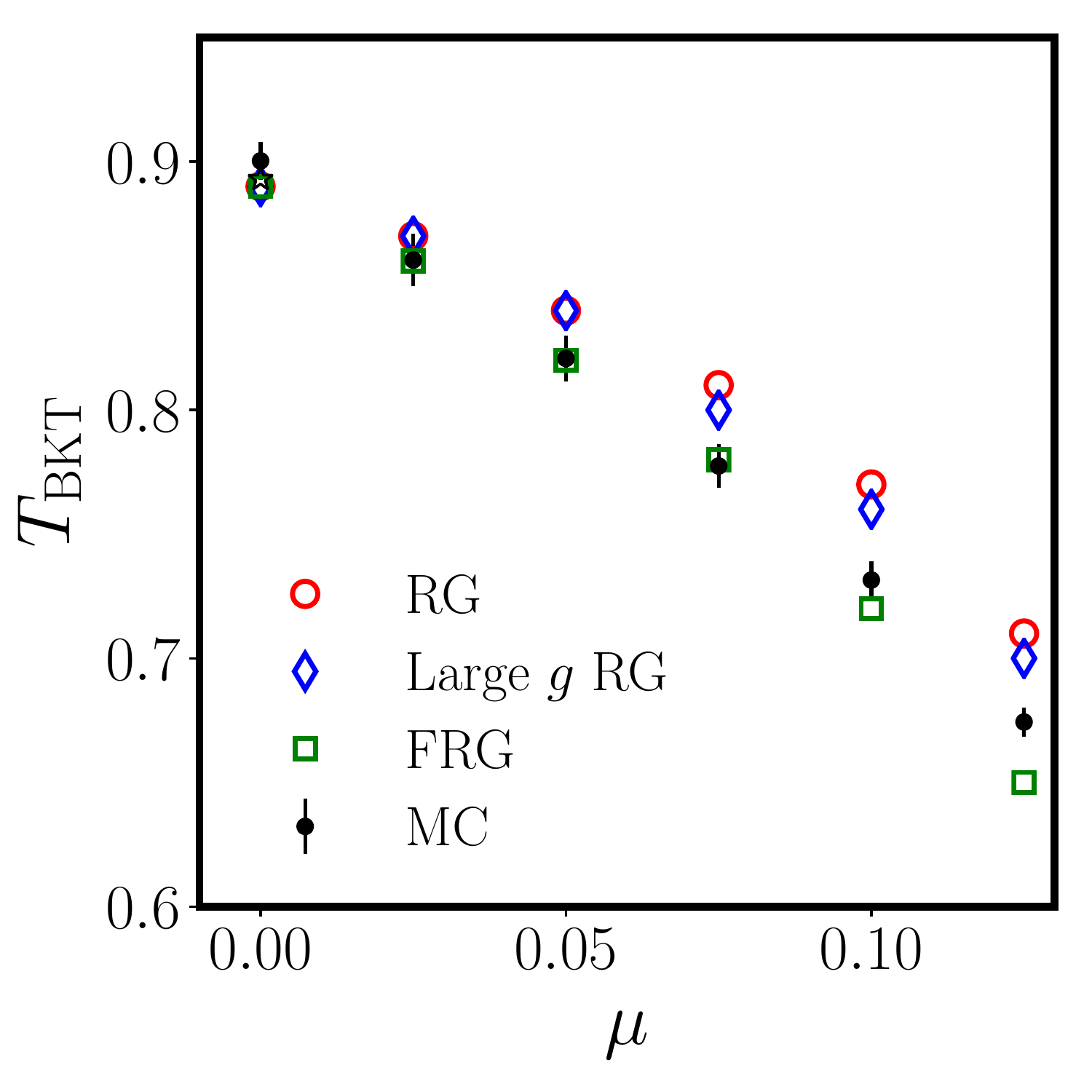}
\caption{\label{Fig5} Critical temperature for the BKT transition in the modified XY model as a function of the additional chemical potential $\mu$. The MC results with finite-size scaling are reported as black points with uncertainties, while circles, diamonds and squares represent different RG schemes described in the text.}
\end{figure}

This continues to work in the modified XY model \eqref{H_2}: as the chemical potential $\mu$ grows, topological fluctuations at increasingly smaller scales enhance the interplay between longitudinal and transverse excitations and seem to undermine the validity of both assumptions (i) and (ii). Yet, the results obtained for the BKT temperature as a function of $\mu$ are in fairly good agreement with the MC results, see Fig.~\ref{Fig5}.

Applying the finite-size scaling procedure described in Ref.~\cite{Olsson1995, Komura2012} to our MC data, we obtain a reliable estimate for the critical temperature $T_{\mathrm{BKT}}$ (black points in Fig.~\ref{Fig5}), which nicely reproduces the high-precision results~\cite{Hasenbusch2005, Komura2012}  at $\mu=0$ (black star). The theoretical estimates for $T_{\mathrm{BKT}}$ obtained by inserting the effective couplings
\begin{align}
J_{\mathrm{eff}}(T)&=J-\frac{T}{4}\,,\label{Jeff2}\\
\mu_{v}^{\mathrm{eff}}(T)&=\gamma^{\mathrm{MC}}J_{\mathrm{eff}}(T) - b_1 \mu +b_2(\mu)T\,,\label{mueff2}
\end{align}
into the initial conditions for the BKT flow Eqs.~\eqref{BKTeq1}~and~\eqref{BKTeq2}, produce a consistent picture for $T_{\mathrm{BKT}}$ up to $\mu\simeq 0.15$, where the nature of the ground state changes and vortex fluctuations become relevant at $T=0$. The results from the two-step approach with the traditional RG equations are shown as red circles in Fig.~\ref{Fig5}.

It is worth noting that the RG analysis performed using the effective couplings Eqs.~\eqref{Jeff2}-\eqref{mueff2} estimated using the coefficients in Tab.~\ref{tab1} greatly improves the accuracy of the two-step approach in the $\mu=0$ case, with respect to the traditional KT initial condition $\mu_{v}=\frac{\pi^{2}}{2}J_{\mathrm{eff}}$.  As the coupling $\mu$ increases, the RG results increasingly deviate from the MC estimates, in agreement with the expectations of larger corrections to the Coulomb gas approximation. At this stage, it is not clear whether these larger deviations are due to an increase in the interplay between longitudinal and transverse phase fluctuations or to higher-order vortex fugacity corrections to the traditional BKT flow equations.

Higher-order terms in the vortex fugacity can be introduced into the BKT flow equations using different approaches. As a first trial, we employed the modified RG equations described in Ref.~\cite{timm_1996} using the effective couplings discussed in Eqs.~\eqref{Jeff2} and \eqref{mueff2}. This analysis slightly improves the accuracy of the predicted critical temperatures especially at high $\mu$ values, as shown with the blue diamonds in Fig.~\ref{Fig5}. To further address the issue of higher-order fugacity corrections to the BKT flow in Eqs.~\eqref{BKTeq1} and \eqref{BKTeq2} in the next section we perform the two-step approach treating the Coulomb gas problem via the functional RG approach.

\subsection{Functional renormalization-group flow for vortex unbinding}

The functional RG approach (FRG) is based on the possibility to write an exact RG equation for the effective action~\cite{Wegner1973,Polchinski1984,Wetterich1993,Morris1994}, which may then
be solved by projecting it onto a restricted theory (coupling) space with a chosen ansatz~\cite{Berges2002, Delamotte:2011jk, Dupuis2020}. The main difference between the FRG scheme and the traditional approach by Wilson lies in the introduction of a momentum-dependent infrared regulator $R_{k}(p)$ to remove the divergence of the propagator close to the critical point and to allow for the derivation of nonperturbative flow equations as the cutoff scale $k$ is lowered. 

In principle, the use of a nonperturbative framework should allow one to study the RG flow for the XY model beyond the quadratic (Villain) limit without the need for the two-step approach described earlier in this section. Nevertheless, efforts to construct an RG flow capable of describing the emergence of topological excitations from microscopic physics were hindered by the difficulty to reproduce the line of fixed points characteristic of the BKT transition~\cite{Jose1977, Graeter1995, Jakubczyk2017} without the use of \emph{ad-hoc} techniques~\cite{Jakubczyk2014}\footnote{Note that despite the difficulties in recovering the correct universal BKT behaviour, the FRG formulation has been able to yield several consistent predictions for the unbinding transition both in the XY model and in 2d Bose gases without the need to explicitly consider vortex fluctuations~\cite{Machado2010,Rancon2012}}.

Therefore, we will perform the FRG study of the unbinding transition within the low-energy sine-Gordon model\,\cite{Nagy2009, Nandori2001}. The flow equations for the vortex fugacity and the stiffness can be conveniently rewritten as
\begin{align}
\label{general_ea_u}
\partial_k u_k &=
\frac1{2\pi} \int_p ~\frac{\partial_k R_k}{u_k}\left(\frac{P_{k}}{\sqrt{P_k^2-u_k^2}}-1\right),\\
\label{general_ea_z}
\partial_k w_k &= \frac1{2\pi}\int_p \partial_k R_k
\biggl(\frac{u_k^2 p^2 (\partial_{p^2}P_k)^2(4P_k^2+u_k^2)}{4(P_k^2-u_k^2)^{7/2}}\nonumber\\
&-\frac{u_k^2P_k(\partial_{p^2}P_k+p^2\partial_{p^2}^2P_k)}{2(P_k^2-u_k^2)^{5/2}} \biggr)
\end{align}
with cutoff $k=1/r=\exp(-\ell)/a$, inverse mass $w_{k}=1/(2 K_{k})$, fugacity $u_{k}=g_{k}/\pi$ and the inverse propagator $P_{k}=w_{k}p^{2}+R_{k}$. The flow Eqs.~\eqref{general_ea_u} and \eqref{general_ea_z} have been derived within the derivative expansion approximation, and their nonperturbative character is apparent from the presence of arbitrary powers of the coupling $u_{k}$. The flow Eqs.~\eqref{general_ea_u} and \eqref{general_ea_z} have been shown to reproduce the salient features of the BKT transition irrespectively of the choice of the regulator function~\cite{Nagy2009,Nandori2001}, and to consistently reproduce the results of Zamolodchikov's $c$-theorem\,\cite{Zomolodchikov1986} in the sine-Gordon model and its generalisations~\cite{Bacso2015,Defenu2019ell}. Moreover, when more advanced functional truncations are considered, the FRG approach yields accurate predictions for the excitation spectrum of the sine-Gordon model~\cite{Daviet2019}.
\begin{figure*}[ht]
\centering
\includegraphics[width=0.8\linewidth]{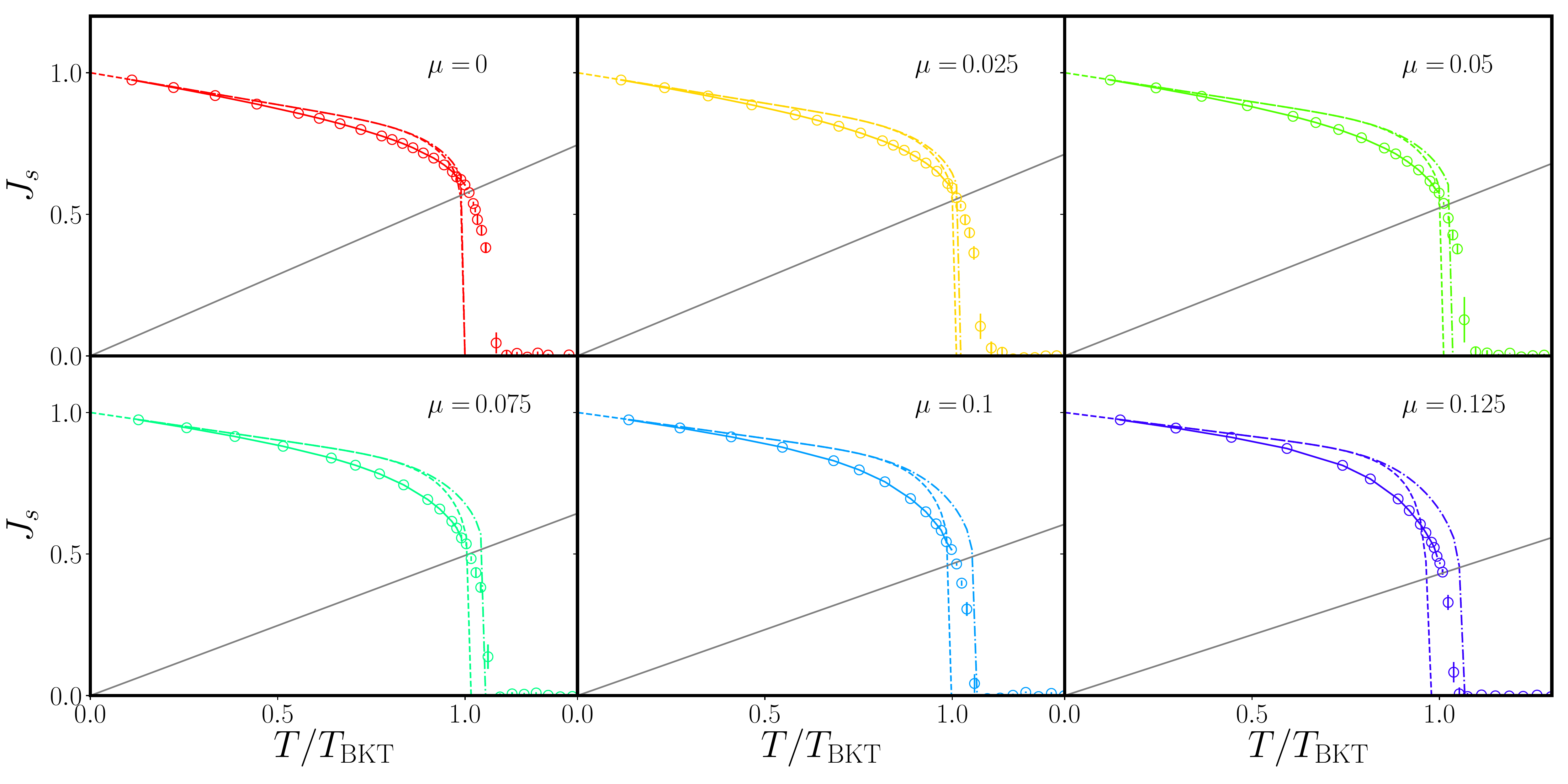}
\caption{\label{Fig6}MC results for the superfluid stiffness $J_{s}$ (circles with error bars) are compared with the predictions from the two-step approach based on the traditional BKT flow Eqs.~\eqref{BKTeq1} and\,\eqref{BKTeq2} (dash-dotted) and the FRG flow Eqs.~\eqref{single_b1_exact} and\,\eqref{single_b1_exact2} (dashed), both with the initial couplings given in Eq.~\eqref{Jeff2} and\,\eqref{mueff2}. From top left to bottom right we show the results for $\mu=0,0.025,0.05,0.075,0.1,0.125$.}
\end{figure*}

For explicit computations it is necessary to specify the form of the regulator function. One straightforward choice is the power-law regulator
\begin{align}
\label{reg_f}
R_{k}(p)/p^{2}=a\left(\frac{k^{2}}{p^{2}}\right)^{b}\,,
\end{align}
where $a$ and $b$ are two free parameters, which are chosen based on an optimisation criterion. However, the common criteria to optimize the regulator function within the FRG approach do not apply to the computation of nonuniversal quantities\,\cite{Canet2003, Litim2000}. Furthermore, the universal features of the BKT transition, such as the jump of the superfluid stiffness, are reproduced by the flow eqs.~\eqref{general_ea_u} and \eqref{general_ea_z} independently from the regulator function $R_{k}$ \cite{Nagy2009} and they cannot be used as a criterion for the choice of the regulator.

Therefore, we are going to use a different criterion for the choice of the optimal values of the parameters $a$ and $b$. To obtain the perturbative BKT flow in Eqs.~\eqref{BKTeq1} and \eqref{BKTeq2}, one
has to assume a short-distance regularization, which traditionally relies
on considering the Coulomb gas charges as hard disks of finite
radius~\cite{KT}. As discussed above, the RG flow produced by such phenomenological regularization has been shown to produce also quantitatively reliable result for small vortex fugacities, see Fig.~\eqref{Fig5}. Then, we choose the parameters $a$ and $b$ in the regulator function \eqref{reg_f} in such a way that the flow Eqs.~\eqref{general_ea_u} and \eqref{general_ea_z} reproduce Eqs.~\eqref{BKTeq1} and \eqref{BKTeq2} at leading order in the vortex fugacity $u_{k}$.

Such optimisation procedure yields the optimal parameter $b=1$, which leads to the analytical flow equations
\begin{align}
\label{single_b1_exact}
(2+k\partial_k)\tilde{u}_k =& \frac{a}{2\pi z_k \tilde{u}_k}  \left[a -  \sqrt{a^{2} - \tilde{u}_k^2} \right] \\
k\partial_k z_k =& -\frac{a}{24\pi} \frac{\tilde{u}_k^2}{[a^{2} - \tilde{u}_k^2]^\frac{3}{2}}
\label{single_b1_exact2}
\end{align}
where $\tilde{u}_{k}=u_{k}/k^{2}$ and $a=1/\sqrt{6\pi^{2}}$ to reproduce the BKT flow at leading order. The estimate of the critical temperatures obtained by the nonperturbative flow Eqs.~\eqref{single_b1_exact} and \eqref{single_b1_exact2} with initial conditions Eq.~\eqref{Jeff2} and \eqref{mueff2} are shown as green squares in Fig.~\ref{Fig5}. We find that the FRG approach produces better results than the traditional BKT flow or the one presented in Ref.~\cite{timm_1996} and always agrees with MC within uncertainties, apart from $\mu=0.125$ very close to the vortex lattice transition.

\subsection{The superfluid stiffness}

Apart from the prediction for the critical temperature $T_{\mathrm{BKT}}$, it is interesting to verify the reliability of the two-step approach in modeling different thermodynamic quantities. The most relevant quantity for our analysis is the superfluid stiffness $J_{s}(T)$, which is obtained in our model by renormalising the low-temperature effective stiffness $J_{\mathrm{eff}}$ in Eq.~\eqref{Jeff}. The results for the superfluid stiffness obtained both by the traditional RG flow and the FRG generalisations are shown in Fig.~\ref{Fig6}. As expected from our optimisation procedure the results of the RG flow eqs.~\eqref{BKTeq1} and \eqref{BKTeq2} are very close to the FRG ones for small vortex fugacities $\mu=0, 0.025, 0.05$, see first three panels in Fig.~\ref{Fig6}. 
As $\mu$ grows, the FRG results for the superfluid show a much better agreement with the MC data than the perturbative RG.

One crucial feature which emerges from the comparison between the FRG and the RG estimates for the superfluid stiffness with respect to the numerical data is that the functional shape of the MC $J_s(T)$ below $T_{\mathrm{BKT}}$ could not be correctly reproduced by any of these approaches. Although the FRG results show the correct unbinding point, they still slightly overestimate the value of $J_s$ below $T_{\mathrm{BKT}}$.
The stronger depletion of the MC $J_s(T)$ approaching $T_{\mathrm{BKT}}$ from below appears to persist in the thermodynamic limit, while above the critical temperature the MC $J_s(T)$ are reduced to reproduce the discontinuous jump in the thermodynamic limit.
In conclusion, the pre-critical descent seems to be a direct indication of the enhanced interplay between spin-wave fluctuations and vortices at high vortex fugacities.

\section{Conclusions}
The main goal of the paper is to prove the possibility to extract accurate nonuniversal quantities ($J_s(T)$, $T_\text{BKT}$) for superfluid systems close to the vortex unbinding transition using the RG description for the Coulomb gas. The procedure to connect the low-energy Coulomb gas description with the microscopic model by effective bare initial conditions in the BKT flow has been widely tested on the square-lattice XY model~\cite{Villain1975, Jose1977, Defenu2017,LBGiamarchiCastellani, Mondal_LB_PRL107, Bighin2016, Janke1993}, but its accuracy for large vortex fugacities had still to be investigated.  In order to prove the feasibility of such a two-step procedure we employed MC simulations to study a modified version of the XY model, where the vortex-core energy can be tuned by a parameter $\mu$ in the Hamiltonian~\eqref{H_2}. The values of the superfluid stiffness ($J_{s}$), the diamagnetic ($J_{d}$) and paramagntic ($J_{p}$) currents, see Fig.~\ref{Fig3},  and the unbinding temperature $T_{\mathrm{BKT}}$ have been numerically derived and compared with the theoretical predictions obtained by inserting the initial conditions given by Eqs.~\eqref{K_0} and \eqref{g_0} with the effective couplings in Eqs.~\eqref{Jeff2} and \eqref{mueff2} into the BKT and FRG flows.

A summary of our analysis can be found in Fig.~\ref{Fig5}, where the MC value of $T_{\mathrm{BKT}}$ is compared with the one obtained by the two-step procedure using different RG approaches. As expected, the reliability of the RG flow in Eqs.~\eqref{BKTeq1} and \eqref{BKTeq2} decreases for lower vortex core energies $\mu_{v}$. This discrepancy in the prediction from the two-step approach may be partially repaired by the introduction of the non-perturbative flow Eqs.~\eqref{single_b1_exact} and \eqref{single_b1_exact2}, which yield accurate estimates for $T_{\mathrm{BKT}}$ in the whole $\mu$ range.

Conversely, the study of the superfluid stiffness reported in Fig.~\ref{Fig6} shows that even the FRG result cannot capture the pronounced depletion in the superfluid stiffness occurring below $T_{\mathrm{BKT}}$, which is most probably the result of the large interplay between transverse and longitudinal fluctuations occurring at large $\mu$. This interplay cannot be captured by the two-step approach, and hence this feature remains unmatched in the RG analysis. 
Surprisingly, these missing features do not preclude accurate estimates of the critical temperatures via the FRG approach.

In conclusion, our analysis proves that the BKT flow equations Eqs.~\eqref{BKTeq1} and \eqref{BKTeq2} represent more than just a low-energy description of the universal behaviour of 2D superfluid systems and may more generally be used to construct a quantitative theory for the unbinding transition once a suitable value for the bare superfluid stiffness $J_{\mathrm{eff}}$ has been identified. For this purpose, we employed the numerical simulations for the vortex density in the system to determine the effective vortex core energy via the low-temperature ansatz in Eq.~\eqref{nu_v_fit}, which, once inserted into the RG flow equations, provides a consistent picture for vortex unbinding.

\textit{Acknoledgements.} We thank J. Lorenzana, N. Dupuis, G. Gori, and A. Trombettoni for stimulating discussions. This work is
supported by the Deutsche Forschungsgemeinschaft (DFG, German Research Foundation) via Project-ID 273811115 - SFB1225 (ISOQUANT) and under Germany’s Excellence Strategy “EXC-2181/1-390900948” (the Heidelberg STRUCTURES Excellence Cluster), by the Italian MAECI under the Italian-India collaborative project SUPERTOP-PGR04879, by
the Italian MIUR project PRIN 2017 No. 2017Z8TS5B,
by Regione Lazio (L. R. 13/08) under project SIMAP, by Sapienza University under project Ateneo 2019 (Grant No. RM11916B56802AFE) and by the G\"{o}ran Gustafsson Foundation for Research in Natural Sciences and Medicine.

\appendix

\section{Numerical Analysis}
\label{sec:MC}
We have performed MC simulations of the Hamiltonian in Eq.~\eqref{H_2}. The linear system sizes considered are $L=8, 16, 32, 64, 128, 256$ with periodic boundary conditions. For each value of $\mu\in\{0,0.025,0.05,0.075,0.1,0.125\}$ and size, we have run $10^4$ MC steps and discarded the transient regime occurring in the first $2\times 10^3$ steps. Each MC step consists of five canonical Metropolis spin flips of the whole lattice, followed by ten micro-canonical overrelaxation sweeps of all spins. 
The thermalization at low temperatures has been sped up by a temperature annealing procedure. Finally, the observables measured have been averaged both over the canonical ensemble (thermal average) and over five independent samples.

The explicit expressions for the diamagnetic ($J_d$) and the paramagnetic ($J_p$) current (along $\hat{x}$) for the model Eq.~\eqref{H_2} studied are
\begin{widetext}
\begin{align}
\label{Jdmicro}
J_d^{{x}}&= \frac{1}{N}\Bigl\langle J \sum_i  \cos(\th_i -\th_{i+x})- 2\mu\sum_i \Big( \cos(\th_i - \th_{i+x})-\cos(\th_{i+x+y} -\th_{i+y}) \Big)^2 \\
         &\quad+2\mu\sum_i I_{P_i} \Big( \sin(\th_i - \th_{i+x}) + \sin(\th_{i+x+y} -\th_{i+y}) \Big)  \Bigr\rangle\,,\notag\\
\label{Jpmicro}
J_p^{{x}}&= \frac{1}{NT} \Big[ \Bigl\langle \Big( J \sum_i \sin(\th_i -\th_{i+x}) - 2\mu \sum_i I_{P_i} \big( \cos(\th_i - \th_{i+x}) -\cos(\th_{i+x+y} -\th_{i+y}) \big) \Big)^2 \Bigr\rangle \\
&\quad-\Bigl\langle  J \sum_i \sin(\th_i -\th_{i+x}) - 2\mu \sum_i I_{P_i} \Big( \cos(\th_i - \th_{i+x}) -\cos(\th_{i+x+y} -\th_{i+y}) \Big) \Bigr\rangle ^2  \Big]\notag
\end{align}
\end{widetext}
with $J_s^{{x}}=J_d^{{x}}-J_p^{{x}}$.

\begin{figure} [!ht]
\includegraphics[width=0.5\textwidth]{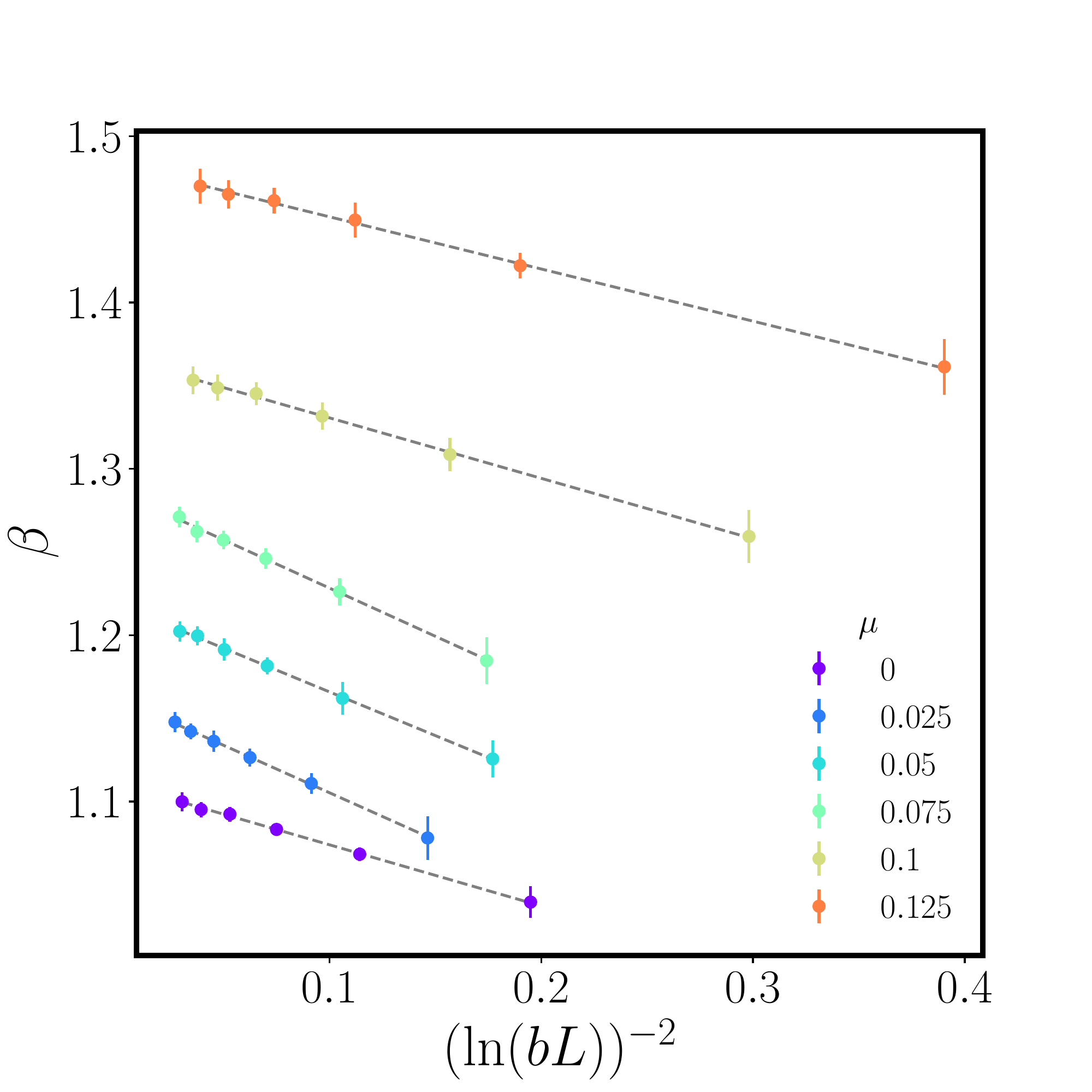}
\caption{Plot of $\beta_{BKT}$ for $L=8, 16, 32, 64, 128, 256$ and $\mu=0, 0.025, 0.05, 0.075, 0.1, 0.125$ from bottom to top. The dashed line is the best fit \eqref{Tfit} used to find $T_{BKT}(\infty)$. The error bars of each $\beta_{BKT}(L, \mu)$ have been computed by both considering the finite spacing dividing the discrete values of the temperatures and by properly propogating the statistical error of $J_s$. }
\label{Fig7}
\end{figure}

For each size $L$ of the system the Nelson criterion~\cite{Nelson} ${J_{s}}/{T_{\mathrm{BKT}}}=\frac{2}{\pi}$ has been applied to calculate the finite-size unbinding transition temperature $T_{\mathrm{BKT}}(L)$. Then, for each value of $\mu$, the thermodynamic critical temperature $T_{BKT}(\infty)$ is extrapolated by means of the finite-size scaling analysis based on the behaviour of the BKT correlation function close to criticality: $\xi \simeq A\, \exp(c/\sqrt{t})$, with $t$ the reduced temperature $t= (T- T_{BKT})/ T_{BKT}$. The relation used to fit our data and extrapolate $T_{BKT}$ is~\cite{Komura2012}
\begin{equation}
  \label{Tfit}
\beta_{BKT}(L)= \beta_{BKT}(\infty)(1- \frac{c^2}{\ln(bL)^2}),
\end{equation}
with $b$ and $c$ fitting paramenters.
The trend of $\beta_{BKT}(L)$ as a function of $(\ln(bL))^{-2}$ is shown for each $\mu$ and for the best fitted parameters in Fig.~\ref{Fig7}.

\section{Vortex Lattice}
\label{app:vlattice}
In Sec.\,IV of the main text, it has already been explained that we expect a transition at $\mu\approx0.15$ between two different zero-temperature ground states: the vortex vacuum with $\rho_{v}(T=0)=0$ for $\mu<\mu_{c}$, and a vortex crystal with $\rho_{v}(T=0)=1$ for $\mu>\mu_{c}$. Even if the study of the vortex crystal phase was not the main scope of the present paper, we would like to present here some results at $\mu=0.15, 0.2$, which have been excluded from the analysis in the main text as they belong to a different critical scenario.
\begin{figure*}[t!]
	\centering
	\subfigure[]{\label{Fig8a}\includegraphics[width=.32\textwidth]{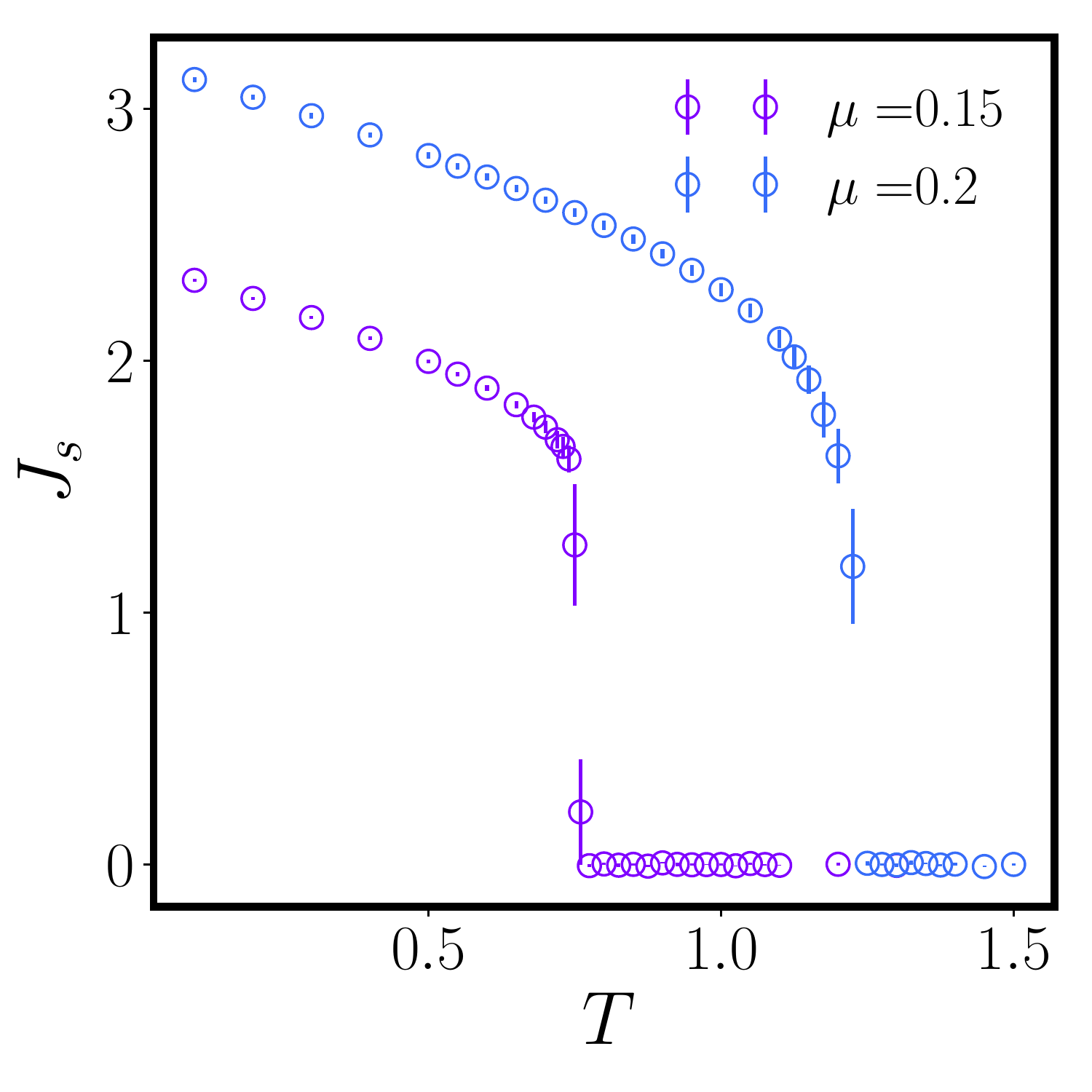}}
	\hfill
	\subfigure[]{\label{Fig8b}\includegraphics[width=.32\textwidth]{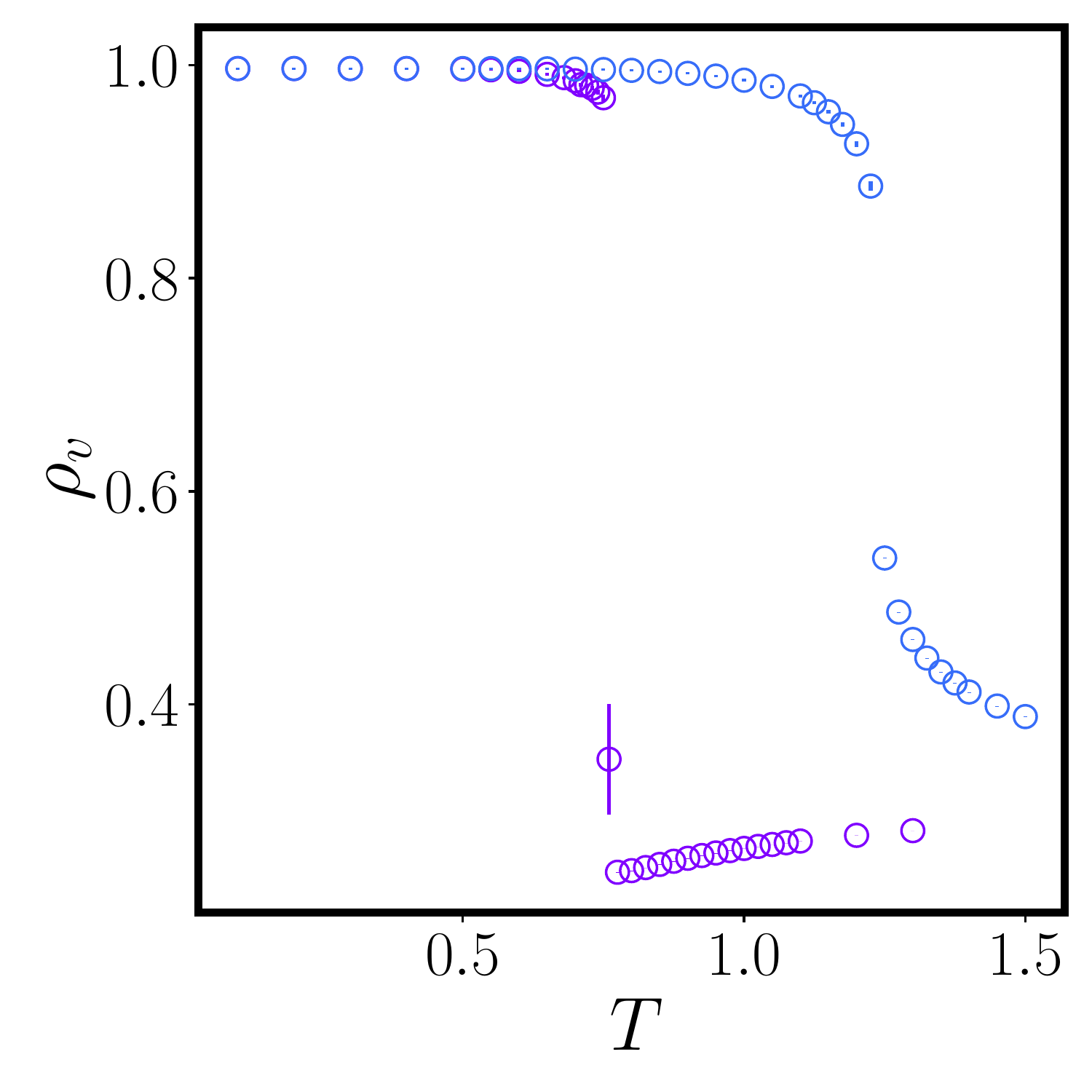}}
	\hfill
	\subfigure[]{\label{Fig8c}\includegraphics[width=.32\textwidth]{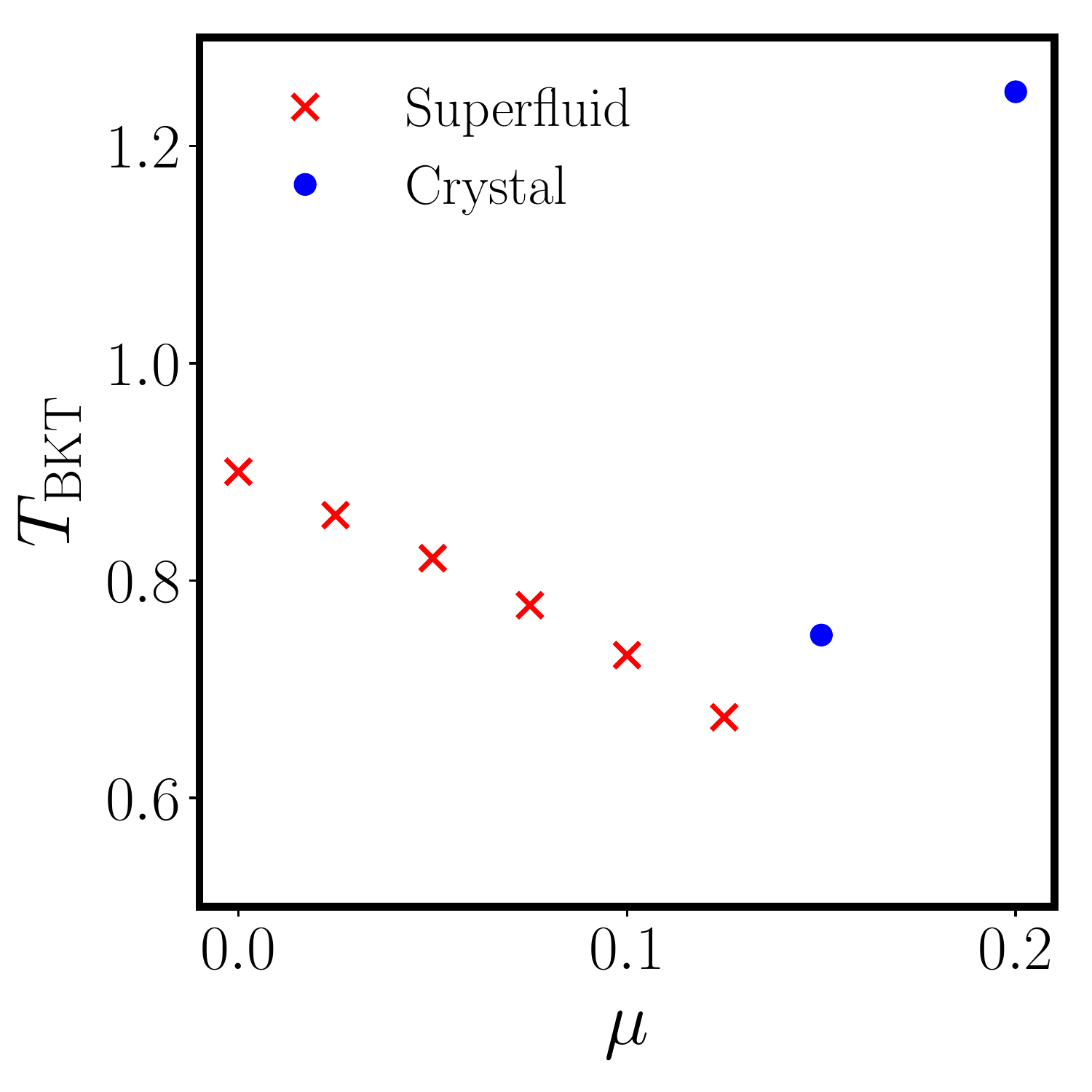}}
	\caption{\label{Fig8} MC analysis of the modified XY model in the stability region of the vortex crystal. Panel (a) reports the superfluid stiffness for $\mu=0.15, 0.2$ (from bottom to top). Panel (b) reports the vortex density for the same values of $\mu$, with larger $\mu$ resulting in larger $\rho_v$. Panel (c) shows the critical temperature of the system both for the vortex unbinding transition (red crosses, data from Fig.~\ref{Fig5}) and for the crystal melting transition (blue circles). }
\end{figure*}

The numerical analysis of the melting transition at $\mu\gtrsim 0.15$ is reported on Fig.~\ref{Fig8}. The superfluid stiffness in Fig.~\ref{Fig8a} displays qualitatively the same behaviour as in the vortex-unbinding regime.
The presence of a different ground state is reflected in the zero-temperature value of the superfluid stiffness which increases with $\mu$, as expected since vortex fluctuation are in this case relevant also at low temperatures.
On the other hand, the superfluid stiffness still presents a sharp jump at the melting point of the 2D crystal.
Even if the melting transition in temperature is expected in two dimensions to belong to the BKT universality class, the present data do not have enough precision to investigate whether this jump is consistent with a BKT transition or rather has the characteristic of a continuous transition.

Nonetheless, more insight can be gained by inspecting the behaviour of the vortex density for the two cases $\mu=0.15,0.2$ shown in Fig.~\ref{Fig8b}. The two curves display a rather different behaviour: in the $\mu=0.15$ case one has $\rho_{v}=1$ at $T\simeq 0$, indicating the crystal phase; as $T$ increases the vortex density drops (almost) discontinuously to $\rho_{v}<0.5$ and, then, slowly increases toward a high-temperature value in agreement with the one observed for the disordered phase at $\mu=0.125$. Conversely, the vortex density at $\mu=0.2$ decreases monotonically (and more smoothly) as a function of the temperature. Partial justification for this behaviour can be found by comparing this behaviour with the one of the lattice Coulomb gas studied in Ref.~\cite{LeeTeitel}. There, the vortex crystal at low temperatures appears for fugacities $g\gtrsim 0.4$ and melts either with a first order transitions for  $g\approx 0.4$ or with a second order phase transition. Our results suggest that the case $\mu=0.15$ belongs to the first scenario, while $\mu=0.2$ to the second one.

The validity of the Coulomb gas description for this model even in the crystal phase is supported by the behaviour of the critical temperature as a function of $\mu$. In the superfluid region $T_{\mathrm{BKT}}$ decreases with increasing $\mu$ until one encounters the critical threshold $\mu_{c}\approx0.15$, above which the critical temperature of the system starts increasing with $\mu$. The resulting curve is nonmonotonic but perfectly continuous, as in the neutral Coulomb gas case. This behaviour is confirmed by Fig.~\ref{Fig8c}, where the critical temperatures for the superfluid and the crystal melting transition are reported vs.~$\mu$. Further numerical analysis would be necessary to clarify possible differences arising in the modified XY model, due to the coupling of vortices and spin waves, with respect to the Coulomb gas case~\cite{LeeTeitel}.  

\bibliography{Biblio_XYmu}

\end{document}